\documentclass{article}%
\usepackage{url}
\usepackage{amsmath}
\usepackage{amssymb}
\usepackage[conf]{ametsoc}
\usepackage{graphicx}
\usepackage{amsfonts}%
\renewcommand{\cite}[2][]{\citep[#1]{#2}}
\begin{document}

\title{An overview of the coupled atmosphere-wildland\\fire model WRF-Fire\thanks{Paper J7.1, 91st American Meterological Society
Annual Meeting, Seattle, WA, January 2011.} }
\author{Jan Mandel \quad Jonathan D. Beezley\\University of Colorado Denver, Denver, CO
\and Adam K. Kochanski\\University of Utah, Salt Lake City, UT}
\maketitle

\begin{center}%
\tableofcontents

\end{center}

\section{Introduction}

Wildland fire is a complicated multiscale process. Fortunately, a practically
important range of wildland fire behavior can be captured by the coupling of a
mesoscale weather model with a simple 2D fire spread model
\cite{Clark-1996-CAF,Clark-1996-CAM}. Weather has a major influence on
wildfire behavior; in particular, wind plays a dominant role in the fire
spread. Conversely, the fire influences the weather through the heat and vapor
fluxes from burning hydrocarbons and evaporation.
The buoyancy created by the heat from the fire can
cause tornadic strength winds, and the wind and the moisture from the fire
affect the atmosphere also away from the fire. It is well known that a large
fire \textquotedblleft creates its own weather.\textquotedblright\ The correct
wildland fire shape and progress result from the two-way interaction between
the fire and the
atmosphere~\cite{Clark-1996-CAF,Clark-1996-CAM,Clark-2004-DCA,Coen-2005-SBE}.

\subsection{Origins and the current state of the model}

WRF-Fire \cite{Mandel-2009-DAW} combines the Weather Research and Forecasting
Model (WRF) \cite{Skamarock-2008-DAR} with a semi-empirical fire spread model,
based on the level-set method. WRF-Fire has grown out of the NCAR's CAWFE code
\cite{Clark-1996-CAF,Clark-1996-CAM,Clark-2004-DCA,Coen-2005-SBE}, which
consists of the Clark-Hall mesoscale atmospheric model coupled with a
tracer-based fire spread model, using the spread rate computed from McArthur's
\cite{Noble-1980-MFM} and later \cite{Rothermel-1972-MMP} formula. Although
the Clark-Hall model has many good properties, it is a legacy serial code, not
supported, and difficult to modify or use with real data, while WRF is a
parallel supported community code routinely used for real runs. See
\cite{Coen-IWF} for a further discussion of their relative merits in the
wildland fire application. \cite{Patton-2004-WCA} proposed a combination of
WRF with the tracer-based model from CAWFE, formulated a~road map, and made
the important observation that the innermost domain of the weather code, which interacts
directly with the fire model, needs to run in the Large Eddy Simulation (LES)
mode. Patton ported the tracer-based code to Fortran 90, rewrote the heat flux
insertion for WRF variables, and produced a prototype code coupled with WRF,
which then served as the foundation of further development. However, instead
of using the existing tracer-based code, the fire module in WRF-Fire was
developed based on the level-set method \cite{Osher-2003-LSM}, among other
reasons, because the level-set function can be manipulated more easily than
tracers for data assimilation. Thus, only the fuel variables and the
subroutine for the calculation of the fire spread rate remained from CAWFE.

Since the paper \cite{Mandel-2009-DAW} was written in 2007, a number of
algorithms and other aspects have changed, new features were added, and
WRF-Fire has been released as a part of WRF starting with version 3.2 in April
2010 \cite{Dudhia-2010-WRF,Wang-2010-AUG}. The latest version of WRF-Fire with
new features and fixes which have not made it into the WRF download yet, plus
additional visualization tools, guides, and diagnostic utilities, are
available from the developers at \url{openwfm.org}.

\subsection{Computational simulations}

While WRF-Fire takes advantage of the CAWFE experience, WRF is quite different
from the Clark-Hall atmospheric model and the fireline propagation algorithm
is also different. Thus, it needs to be demonstrated that WRF-Fire can deliver
similar results as CAWFE, and WRF-Fire needs to be validated against real
fires. \cite{Kim-2011-RDE} has shown that the level set method in the fire
module can advect the fire shape correctly, just like the CAWFE tracer code in
\cite{Clark-2004-DCA}. \cite{Jenkins-2010-FDF} studied the role of wind
profile on fire propagation speed and the shape of the fireline and
demonstrated fireline fingering behavior on ideal examples, as in
\cite{Clark-1996-CAF,Clark-1996-CAM}. \cite{Kochanski-2010-EFP} compared
simulation results with measurements on the FireFlux grass fire experiment
\cite{Clements-2007-ODW}. \cite{Dobrinkova-2010-WAB} simulated a fire in
Bulgarian mountains using real meteorological and geographical data, and ideal
fuel data. \cite{Beezley-2010-SMC} simulated a fire in Colorado mountains
using real data from online sources. A mesoscale simulation can run faster
than real time on a small cluster \cite{Jordanov-2011-SWF}.

\subsection{Related work}

Wildland fire models range from tools based based on fire spread rate formulas
\cite{Rothermel-1972-MMP,Rothermel-1983-HTP}, such as BehavePlus
\cite{Andrews-2007-BFM} and FARSITE \cite{Finney-1998-FFA}, suitable for
operational forecasting, to sophisticated 3D computational fluid dynamics and
combustion simulations suitable for research and reanalysis, such as FIRETEC
\cite{Linn-2002-SWB} and WFDS \cite{Mell-2007-PAM}. BehavePlus, the PC-based
successor of the calculator-based BEHAVE, determines the fire spread rate at a
single point from fuel and environmental data; FARSITE uses the fire spread
rate to provide a 2D simulation on a PC; while FIRETEC and WFDS require a
parallel supercomputer and run much slower than real time.

The level set-method was used for a surface fire spread model in
\cite{Mallet-2009-MWF}. \cite{Filippi-2009-CAF} coupled the atmospheric model,
Meso-nh, with fire propagation by tracers. Tiger \cite{Mazzoleni-T2F} uses a
2D combustion model based on reaction-convection-diffusion equations and a
convection model to emulate the effect of the fire on the wind. FIRESTAR
\cite{Morvan-2004-MPW} is a physically accurate wildland fire model in two
dimensions, one horizontal and one vertical. UU LES-Fire \cite{Sun-2009-IFC}
couples the University of Utah's Large Eddy Simulator with the tracer-based
code from CAWFE. See the survey by \cite{Sullivan-2009-RWF} for a number of
other models.

\section{Physical fire model and fuels}

\label{sec:physical-model}

The physical model consists of subroutines computing the spread rate and the
burn rate, and it is essentially the same as a subset of CAWFE
\cite{Clark-1996-CAF,Clark-1996-CAM,Clark-2004-DCA,Coen-2005-SBE},
\cite{Rothermel-1972-MMP}, and\ BEHAVE \cite{Andrews-2007-BFM}. 

\subsection{Fuel properties}

Fuel is characterized by a vector of quantities, which are given at every
point of the domain. To simplify the specification of fuel properties, fuels
are given as one of 13 \cite{Anderson-1982-ADF} categories, which are preset
vectors of values of the fuel properties above. These values are specified in
an input text file, and they can be modified by the user. The user can also
specify completely new, custom fuels.

\subsection{Fire spread rate}

Mathematically, the fire model is posed in the horizontal $(x,y)$ plane. The
semi-empirical approach to fire propagation used here assumes that the fire
spread rate is given by the modified Rothermel's formula%
\begin{equation}
S=R_{0}\left(  1+\phi_{W}+\phi_{S}\right)  . \label{eq:spread}%
\end{equation}
The spread rate depends on fuel properties, the wind speed $U$, and the
terrain slope $\tan\phi$, exactly as in \cite{Rothermel-1972-MMP}, except that
some of the input quantities are metric so they are first converted from
metric to English units (lb-ft-min) to avoid changing the numerous constants
in the computation from \cite{Rothermel-1972-MMP}; the wind is limited to
between $0$ to $30$m/s; the slope is limited to nonnegative values; the fuel
mass with moisture is given rather than dry fuel as in
\cite{Rothermel-1972-MMP}; and a new category is introduced for chaparral from
\cite[eq. (1)]{Coen-2001-CAM}. In either case, the spread rate can be written
as%
\begin{equation}
S=\max\left\{  S_{0},R_{0}+c\min\left\{  e,\max\left\{  0,U\right\}  \right\}
^{b}+d\max\left\{  0,\tan\phi\right\}  \right\}  , \label{eq:spread-WRF-Fire}%
\end{equation}
where $S_{0},R_{0}$,$b,c,d,e$ are fuel-dependent coefficients, which is how
the spread rate is represented in WRF-Fire internally. These coefficients are
stored for every grid point. At a fixed point on the fireline, denote by
$\vec{n}$ the outside normal to the fire region, $\overrightarrow{U}$ the wind
vector, and $z$ the terrain height. The normal component of the wind vector,
$U=\overrightarrow{U}\cdot\vec{n}$, and the normal component of the terrain
gradient, $\tan\phi=\nabla z\cdot\vec{n}$, are used to determine the spread
rate, which is interpreted as the spread rate in the normal direction $\vec
{n}$.

\subsection{Fuel burned and heat released}

Each location starts with fuel fraction $F=1$. Once the fuel is ignited at a
time $t_{i}$, the fuel fraction decreases exponentially,%
\begin{equation}
F\left(  t\right)  =\exp\left(  -\frac{\left(  {t-t}_{i}\right)  }{T_{f}%
}\right)  ,\quad t>{t_{i}} \label{eq:fuel}%
\end{equation}
where $t$ is the time, $t_{i}$ is the ignition time, $F_{0}$ is the initial
amount of fuel, and $T_{f}$ is the time constant of fuel, i.e., the number of
seconds for the fuel to burn down to $1/e\approx0.3689$ of the original
quantity. Since the fuel burns down to $0.6$ of the original quantity in
$600$s when $w=1000$, we have%
\[
0.6^{\frac{\left(  t-t_{i}\right)  }{600}\frac{1000}{w}}=\exp\left(
-\frac{\left(  {t-t}_{i}\right)  }{T_{f}}\right)  ,
\]
which gives%
\[
T_{f}=-\frac{600}{1000\ln0.6}\approx\frac{w}{0.8514}.
\]
The input coefficient $w$ is used in WRF-Fire rather than $T_{f}$ for
compatibility with existing fuel models and literature

The average sensible heat flux density released in time interval $\left(
t,t+\Delta t\right)  $ is computed as%
\begin{equation}
\phi_{h}=\frac{F\left(  t\right)  -F\left(  t+\Delta t\right)  }{\Delta
t}\frac{1}{1+M_{f}}w_{\ell}h\quad\text{(J/m}^{2}\text{/s)}
\label{eq:sensible-heat-flux}%
\end{equation}
and the average latent heat (i.e., moisture) flux density is given by%
\begin{equation}
\phi_{q}=\frac{F\left(  t\right)  -F\left(  t+\Delta t\right)  }{\Delta
t}\frac{M_{f}+0.56}{1+M_{f}}Lw_{\ell}\quad\text{(J/m}^{2}\text{/s)}
\label{eq:latent-heat-flux}%
\end{equation}
where 0.56 is the estimated mass ratio of the water output from combustion to
the dry fuel, and $L=2.5\cdot10^{6}$ J/kg is the specific latent heat of
condensation of water at $0$ $^{o}$C, used for nominal conversion of moisture
to heat. This computation is from CAWFE.

\section{Mathematical core of the fire model}

\label{sec:core}

The model maintains level set function $\psi$, time of ignition $t_{i}$, and
the fuel fraction $F$. These quantities are all represented by their values at
the centers of the fire mesh cells.

\subsection{Fire propagation by the level set method}

\label{sec:levelset}

This section follows \cite{Mandel-2009-DAW}. Denote point on the surface by
$\mathbf{x}=\left(  x,y\right)  $. The burning region at time $t$ is
represented by a level set function $\psi=\psi\left(  \mathbf{x},t\right)  $
as the set of all points $\mathbf{x}$ such that where $\psi\left(
\mathbf{x},t\right)  \leq0$. There is no fire at $\mathbf{x}$ if $\psi\left(
\mathbf{x},t\right)  >0$. The fireline is the set of all points $\mathbf{x}$
such that $\psi\left(  \mathbf{x},t\right)  =0$. Since on the fireline, the
tangential component of the gradient $\nabla\psi$ is zero, the outside normal
vector at the fireline is
\begin{equation}
\mathbf{n}=\frac{\nabla\psi}{\left\Vert \nabla\psi\right\Vert }.
\label{eq:normal}%
\end{equation}
Now consider a point $\mathbf{x}\left(  t\right)  $ that moves with the
fireline. Then the fire spread rate $S$ at $\mathbf{x}$ in the direction of
the normal $\mathbf{n}$ is%
\begin{equation}
S=\overrightarrow{n}\cdot\frac{\partial\mathbf{x}}{\partial t}.
\label{eq:normal-spread}%
\end{equation}
From $\psi\left(  \mathbf{x}\left(  t\right)  ,t\right)  =0$ and chain rule,%
\begin{equation}
0=\frac{d}{dt}\psi\left(  x,y,t\right)  =\frac{\partial\psi}{\partial t}%
+\frac{\partial\psi}{\partial x}\frac{\partial x}{\partial t}+\frac
{\partial\psi}{\partial y}\frac{\partial y}{\partial t}=\frac{\partial\psi
}{\partial t}+\left\Vert \nabla\psi\right\Vert \left(  \mathbf{n}\cdot
\frac{\partial\mathbf{x}}{\partial t}\right)  =\frac{\partial\psi}{\partial
t}+S\left\Vert \nabla\psi\right\Vert , \label{eq:chain-rule-fireline}%
\end{equation}
so, from (\ref{eq:normal-spread})\ and (\ref{eq:chain-rule-fireline}), the
level set function is governed by the partial differential equation%
\begin{equation}
\frac{\partial\psi}{\partial t}+S\left(  x\right)  \left\Vert \nabla
\psi\right\Vert =0, \label{eq:levelset}%
\end{equation}
called the level set equation \cite{Osher-2003-LSM}. The spread rate $S$ is
evaluated from (\ref{eq:spread-WRF-Fire}) everywhere on the domain. Since
$S\geq0$, the level set function does not increase with time, and the fire
area cannot decrease, which also helps with numerical stability by eliminating
oscillations of $\psi$ in time.

The level set equation is discretized on a rectangular grid with spacing
$\left(  \triangle x,\triangle y\right)  $, called the fire grid. The level
set function $\psi$ and the ignition time $t_{i}$ are represented by their
values at the centers of the fire grid cells. This is consistent with the fuel
data given in the center of each cell also.

To advance the fire region in time, we use Heun's method (Runge-Kutta method
of order $2$),
\begin{align}
\psi^{n+1/2}  &  =\psi^{n}+\Delta tF\left(  \psi^{n}\right) \nonumber\\
\psi^{n+1}  &  =\psi^{n}+\Delta t\left(  \frac{1}{2}F\left(  \psi^{n}\right)
+\frac{1}{2}F\left(  \psi^{n+1/2}\right)  \right)  , \label{eq:Heun}%
\end{align}
The right-hand side $F$ is a discretization of the term $-S\left\Vert
\nabla\psi\right\Vert $ with upwinding and artificial viscosity,
specifically,
\[
F\left(  \psi\right)  =-S\left(  \overrightarrow{v}\cdot\overrightarrow
{n},\nabla z\cdot\overrightarrow{n}\right)  \left\Vert \overline{\nabla}%
\psi\right\Vert +\varepsilon\widetilde{\triangle}\psi,
\]
where $\overrightarrow{n}=\nabla\psi/\Vert\nabla\psi\Vert$ is computed by
central differences and $\overline{\nabla}\psi=\ \left[  \overline{\nabla}%
_{x}\psi,\overline{\nabla}_{y}\psi\right]  $ is the upwinded finite difference
approximation of $\nabla\psi$ by Godunov's method \cite[p. 58]{Osher-2003-LSM}%
,
\begin{equation}
\overline{\nabla}_{x}\psi=\left\{
\begin{array}
[c]{c}%
\overline{\nabla}_{x}^{+}\psi\text{ if }\overline{\nabla}_{x}^{-}\psi
\leq0\text{ and }\overline{\nabla}_{x}^{+}\psi\leq0,\\
\overline{\nabla}_{x}^{-}\psi\text{ if }\overline{\nabla}_{x}^{-}\psi
\geq0\text{ and }\overline{\nabla}_{x}^{+}\psi\geq0,\\
0\text{ if }\overline{\nabla}_{x}^{-}\psi\leq0\text{ and }\overline{\nabla
}_{x}^{+}\psi\geq0,\\
\text{otherwise }\overline{\nabla}_{x}^{-}\psi\text{ if }\left\vert
\overline{\nabla}_{x}^{-}\psi\right\vert \geq\left\vert \overline{\nabla}%
_{x}^{+}\psi\right\vert ,\\
\quad\quad\quad\quad\quad\overline{\nabla}_{x}^{+}\psi\text{ if }\left\vert
\overline{\nabla}_{x}^{-}\psi\right\vert \leq\left\vert \overline{\nabla}%
_{x}^{+}\psi\right\vert ,
\end{array}
\right.  \label{eq:Godunov}%
\end{equation}
where $\nabla_{x}^{+}\psi$ and $\nabla_{x}^{-}\psi$ are the right and left
one-sided differences%
\begin{align*}
\nabla_{x}^{+}\psi\left(  x,y\right)   &  =\frac{\psi\left(  x+\triangle
x,y\right)  -\psi\left(  x,y\right)  }{\triangle x},\\
\nabla_{x}^{-}\psi\left(  x,y\right)   &  =\frac{\psi\left(  x,y\right)
-\psi\left(  x-\triangle x,y\right)  }{\triangle x},
\end{align*}
and similarly for $\nabla_{y}^{+}\psi$ and $\nabla_{y}^{-}\psi$. Further,
\[
\nabla\psi=\left[  \frac{\nabla_{x}^{+}\psi+\nabla_{x}^{-}\psi}{2}%
,\frac{\nabla_{y}^{+}\psi+\nabla_{y}^{-}\psi}{2}\right]
\]
is the gradient by central differences, $\varepsilon$ is the scale-free
artificial viscosity ($\varepsilon=0.4$ in the computations here), and%
\[
\widetilde{\triangle}\psi=\nabla_{x}^{+}\psi-\nabla_{x}^{-}\psi+\nabla_{y}%
^{+}\psi-\nabla_{y}^{-}\psi
\]
is the scaled five-point Laplacian of $\psi$.

A numerically stable scheme that includes upwinding, such as Godunov's method
(\ref{eq:Godunov}), is required to compute $\left\Vert \nabla\psi\right\Vert $
in the level set equation (\ref{eq:levelset}). However, it seems better to use
standard central differences for $\nabla\psi$ in the computation of the normal
$\mathbf{n}$ in (\ref{eq:normal}), which is needed to evaluate the normal
component of the wind and the slope in (\ref{eq:spread-WRF-Fire}).

Before computing the one-sided differences up to the boundary, the level set
function is extrapolated to one layer of nodes beyond the boundary. However,
the extrapolation is not allowed to decrease the value of the level set
function under the value at either of the points extrapolated from. For
example, when $\left(  i,j\right)  $ is the last node in the domain in the
direction $x,$ the extrapolation
\[
\psi_{i+1,j}=\max\left\{  \psi_{ij}+\left(  \psi_{ij}-\psi_{i-1,j}\right)
,\psi_{ij},\psi_{i-1,j}\right\}  ,
\]
is used, and similarly in the other cases. This modification of the finite
difference method serves to avoid numerical instabilities at the boundary. The
extrapolation at the boundary effectively implements a free boundary
condition. Without the stabilization, a decrease of $\psi$ at a boundary node,
which often happens for nonhomogeneous fuels in real data, is amplified by
extrapolation and because of upwinding, $\psi$ keeps decreasing at that
boundary node forever and developing a large negative spike.

The model does not support fire crossing the boundary of the domain. When
$\psi<0$ is detected near the boundary, the simulation terminates. This is not
a limitation in practice, because the fire should be well inside the domain
anyway for a proper response of the atmosphere.

The ignition time $t_{i}$ in the strip that the fire has moved over in one
timestep is computed by linear interpolation from the level set function.
Suppose that the point $\mathbf{x}$ is not burning at time $t$ but is burning
at time $t+\triangle t$, that is, $\psi\left(  \mathbf{x},t\right)  >0$ and
$\psi\left(  \mathbf{x},t+\triangle t\right)  \leq0$. The ignition time at
$\mathbf{x}$ satisfies $\psi\left(  \mathbf{x},t_{i}\left(  \mathbf{x}\right)
\right)  =0$. Approximating $\psi$ by a linear function in time, we have%
\[
\frac{\psi\left(  \mathbf{x},t_{i}\right)  -\psi\left(  \mathbf{x},t\right)
}{t_{i}\left(  \mathbf{x}\right)  -t}\approx\frac{\psi\left(  \mathbf{x}%
,t+\triangle t\right)  -\psi\left(  \mathbf{x},t_{i}\right)  }{t+\triangle
t-t_{i}\left(  \mathbf{x}\right)  },
\]
and we take%
\begin{equation}
t_{i}(\mathbf{x})=t+\frac{\psi\left(  \mathbf{x},t\right)  \triangle t}%
{\psi\left(  \mathbf{x},t\right)  -\psi\left(  \mathbf{x},t+\triangle
t\right)  }. \label{eq:set-ignition-time}%
\end{equation}

\subsection{Computation of the fuel fraction}

\label{sec:fuel-fraction}

The fuel fraction is approximated over each fire mesh cell $C$ by integrating
(\ref{eq:fuel}) over the fire region. Hence, the fuel fraction remaining in
cell $C$ at time $t$ is given by%
\begin{equation}
1-\frac{1}{\operatorname*{area}\left(  C\right)  }{\iint\limits_{%
\genfrac{}{}{0pt}{}{\mathbf{x}\in C}{\psi\left(  \mathbf{x},t\right)  \leq0}%
}}1-\exp\left(  -\frac{t-t_{i}\left(  \mathbf{x}\right)  }{T_{f}(\mathbf{x}%
)}\right)  d\mathbf{x.} \label{eq:fuel_left}%
\end{equation}
Once the fuel fraction is known, the heat fluxes are computed from
(\ref{eq:sensible-heat-flux}) and (\ref{eq:latent-heat-flux}). This scheme has
the advantage that the total heat released in the atmosphere after the fuel
has completely burned is accurate, regardless of approximations in the
computation of the integral (\ref{eq:fuel_left}). We are looking for a scheme
that is second order accurate when the whole cell is on fire, exact when no
part of the cell $C$ is on fire (namely, returning the value one), and
provides a natural transition between these two cases. Just like the standard
numerical schemes are exact for polynomials of a certain degree, the guiding
principle here is that the scheme should be exact in a collection of
(nontrivial) special cases.

While the fuel time $T_{f}$ can be interpolated as constant over the whole
cell, the level set function $\psi$ and the ignition time $t_{i}$ must be
interpolated more accurately to allow submesh representation of the burning
area and gradual release of the heat as the fireline moves over the cell.
Then, to compute the integral in (\ref{eq:fuel_left}), the cell $C$ is split
into 4 subcells $C_{j}$, and%
\begin{equation}
{\iint\limits_{%
\genfrac{}{}{0pt}{}{\mathbf{x}\in C}{\psi\left(  \mathbf{x},t\right)  \leq0}%
}1-}\exp\left(  -\frac{t-t_{i}\left(  \mathbf{x}\right)  }{T_{f}(\mathbf{x}%
)}\right)  d\mathbf{x=}\sum_{j=1}^{4}{\iint\limits_{%
\genfrac{}{}{0pt}{}{\mathbf{x}\in C_{j}}{\psi\left(  \mathbf{x},t\right)
\leq0}%
}1-}\exp\left(  -\frac{t-t_{i}\left(  \mathbf{x}\right)  }{T_{f}(\mathbf{x}%
)}\right)  d\mathbf{x.} \label{eq:fuel-left-subcells}%
\end{equation}
The level-set function $\psi$ is interpolated bilinearly to the vertices of
the subcells $C_{j}$, and $T_{f}$ is constant on each $C_{j}$. When the whole
cell $C$ is on fire (that is, $\psi\leq0$ on all four vertices of $C$),
$t_{i}$ is interpolated also linearly to the vertices of the subcells $C_{j}$.
However, the case when the fireline crosses the cell $C$ requires a special
treatment of the ignition time $t_{i}$; $t_{i}\left(  \mathbf{x}\right)  $ has
meaningful value only when the node $\mathbf{x}$ is on nodes on fire,
$\psi\left(  \mathbf{x}\right)  \leq0$, and $t_{i}\left(  \mathbf{x}\right)
=0$ on the fireline, i.e., when $\psi\left(  \mathbf{x}\right)  =0$.
Approximating both $\psi$ and $t_{i}$ in the fire region by a linear function
suggests interpolating from the relation%
\begin{equation}
t_{i}-t=c\psi, \label{eq:ignition-prop}%
\end{equation}
for some $c$. We interpolate on the grid lines between two nodes first. If
both nodes are on fire, we interpolate $t_{i}$ bilinearly as before. However,
when one cell center is on fire and one not, say $\psi\left(  \mathbf{a}%
_{1}\right)  >0$, $\psi\left(  \mathbf{a}_{2}\right)  <0$,. we find the
proportionality constant $c$ in (\ref{eq:ignition-prop}) from $t_{i}\left(
\mathbf{a}_{2}\right)  =c\psi\left(  \mathbf{a}_{2}\right)  $, and set
$t_{i}\left(  \mathbf{b}\right)  =c\psi\left(  \mathbf{b}\right)  $ at the
midpoint $\mathbf{b}=\left(  \mathbf{a}_{1}+\mathbf{a}_{2}\right)  /2$. In the
case of interpolation to the node $\mathbf{c=}\left(  \mathbf{a}%
_{1}+\mathbf{a}_{2}+\mathbf{a}_{3}+\mathbf{a}_{4}\right)  /4$ between nodes
$\mathbf{a}_{1},\mathbf{a}_{2},\mathbf{a}_{3},\mathbf{a}_{4}$, we find the
proportionality constant $c$ by solving the least squares problem%
\[
\sum_{\substack{j=1\\\psi\left(  \mathbf{a}_{j}\right)  \leq0}}^{4}\left\vert
t_{i}\left(  \mathbf{a}_{j}\right)  -t-c\psi\left(  \mathbf{a}_{j}\right)
\right\vert ^{2}\rightarrow\min
\]
and set again $t_{i}\left(  \mathbf{c}\right)  =c\psi\left(  \mathbf{c}%
\right)  $.

To compute the integral over a subcell $C_{j}$, we first estimate the fraction
of the subcell that is burning, by
\begin{equation}
\frac{\operatorname*{area}\left\{  (x,y)\in C_{j}:\psi\left(  x,y,t\right)
\leq0\right\}  }{\operatorname*{area}(C_{j})}\approx\alpha=\frac{1}{2}\left(
1-\frac{\sum_{k=1}^{4}\psi\left(  \mathbf{x}_{k}\right)  }{\sum_{k=1}%
^{4}\left\vert \psi\left(  \mathbf{x}_{k}\right)  \right\vert }\right)  ,
\label{eq:approx-area}%
\end{equation}
where $\mathbf{x}_{k}$ are the the corners of the subcell $C_{j}$. The
approximation is exact when no part of the subcell $C_{j}$, is on fire, that
is, all $\psi\left(  \mathbf{x}_{k}\right)  \geq0$ and at least one
$\psi\left(  \mathbf{x}_{k}\right)  >0$; the whole $C_{j}$ is on fire, that
is, all $\psi\left(  \mathbf{x}_{k}\right)  \leq0$ and at least one
$\psi\left(  \mathbf{x}_{k}\right)  <0$; or the values $\psi\left(
\mathbf{x}_{k}\right)  $ define a linear function and the fireline crosses the
subcell diagonally or it is aligned with one of the coordinate directions.

Next, replace $t\left(  \mathbf{x}_{k}\right)  $ by $t$ when $\psi\left(
\mathbf{x}_{k}\right)  >0$ (i.e., the node $\mathbf{x}_{k}$ is not on fire),
and compute the approximate fraction of the fuel burned as
\begin{equation}
\alpha\left(  1-\exp\left(  -\frac{1}{4}\sum_{k=1}^{4}\frac{t_{i}\left(
\mathbf{x}_{k}\right)  -t}{T_{f}}\right)  \right)  \label{eq:approx-fraction}%
\end{equation}

This computation is a second-order quadrature formula when the whole cell is
burning; it is exact when no part of the cell is burning; and it provides a
natural transition between the two. Also, the calculation is accurate
asymptotically when the fuel burns slowly and the approximation $\beta$ of the
burning area is exact.

Optionally, the fuel fraction remaining is computed first by approximating
$\psi$ and $t_{i}$ and by linear functions using finite differences and then
integrating analytically, first in the direction orthogonal to the fireline
and then parallel to the fireline, thus splitting the fire region in the
subcell into trapezoids. This method is exact when $\psi$ and $t_{i}$ are
linear functions, but it is more expensive. Accurate calculation of the fuel
fraction and thus of the heat flux will be important for the simulation of
crown fire, which is ignited when the ground heat flux exceeds a certain
threshold value \cite{Clark-1996-CAM}).

\subsection{Ignition}

\label{sec:ignition}

Typically, a fire starts much smaller than the fire mesh cell size, and both
point and line ignition need to be supported. The previous ignition mechanism
\cite{Mandel-2009-DAW} ignited everything within a given distance from the
ignition line at once. This distance was required to be at least one or two
mesh steps, so that the initial fire is visible on the fire mesh, and the fire
propagation algorithm from Sec.~\ref{sec:levelset} can catch on. This caused
an unrealistically large initial heat flux and an accelerated ignition.

The current ignition scheme achieves submesh resolution and zero-size
ignition. A small initial fire is superimposed on the regular propagation
mechanism, which then takes over. Drip-torch ignition is implemented as a
collection of short ignition segments that grow at one end every time step.
Multiple ignitions are supported.

The model is initialized with no fire by choosing the level set function
$\psi\left(  \mathbf{x},t_{0}\right)  =\operatorname*{const}>0$. Consider an
initial fire that starts at time $t_{g}$ on a segment $\overline{\mathbf{ab}}$
and propagates in all directions with an initial spread rate $S_{g}$ until
distance $r_{g}$ is reached. At the beginning of every time step $t$ such that
$t_{g}\leq t\leq t_{g}+r_{g}/S_{g}$, we construct the level-set function of
the initial fire,%
\begin{equation}
\psi_{i}\left(  \mathbf{x},t\right)  =\operatorname*{dist}\left(
\mathbf{x},\overline{\mathbf{ab}}\right)  -S_{g}\left(  t-t_{g}\right)
\label{eq:ignition-lfn}%
\end{equation}
and replace the level-set function of the model by
\begin{equation}
\psi\left(  \mathbf{x},t\right)  :=\min\left\{  \psi\left(  \mathbf{x}%
,t\right)  ,\psi_{i}\left(  \mathbf{x},t\right)  \right\}  .
\label{eq:ignition-min}%
\end{equation}
For a drip-torch ignition starting from point $\mathbf{a}$ at time $t_{g}$ at
velocity $\mathbf{v}$ until time $t_{h}$, the ignition line at time $t$ is the
segment $\overline{\mathbf{a},\mathbf{a}+\mathbf{v}\left(  \min\left\{
t,t_{h}\right\}  -t_{g}\right)  }$, and (\ref{eq:ignition-lfn}) becomes%
\[
\psi_{i}\left(  \mathbf{x},t\right)  =\operatorname*{dist}\left(
\mathbf{x},\overline{\mathbf{a},\mathbf{a}+\mathbf{v}\left(  \min\left\{
t,t_{h}\right\}  -t_{g}\right)  }\right)  -\min\left\{  r_{g},S_{g}\left(
t-t_{g}\right)  \right\}
\]
followed again by (\ref{eq:ignition-min}), at the beginning of every time step
$t$ such that%
\[
t_{g}\leq t\leq t_{h}+r_{g}/S_{g}.
\]
The ignition time on newly ignited nodes is set to the arrival time of the
fire at the spread rate $S_{g}$ from the nearest point on the ignition segment.

\section{Atmospheric model}

\textbf{\label{sec:WRF}}We summarize some background information about WRF
from \cite{Skamarock-2008-DAR} to the extend needed to understand the coupling
with the fire module.

\subsection{Variables and equations}

The model is formulated in terms of the hydrostatic pressure vertical
coordinate $\eta$, scaled and shifted so that $\eta=1$ at the Earth surface
and $\eta=0$ at the top of the domain. The governing equations are a system of
partial differential equations of the form
\begin{equation}
\frac{d\Phi}{dt}=R\left(  \Phi\right)  , \label{eq:WRF}%
\end{equation}
where $\Phi=\left(  U,V,W,\phi^{\prime},\Theta,\mu^{\prime},Q_{m}\right)  $.
The fundamental WRF variables are $\mu=\mu\left(  x,y\right)  $, the
hydrostatic component of the pressure differential of dry air between the
surface and the top of the domain, written in perturbation form $\mu
=\overline{\mu}+\mu^{\prime}$, where $\overline{\mu}$ is a reference value in
hydrostatic balance; $U=\mu u$, where $u=u\left(  x,y,\eta\right)  $ is the
Cartesian component of the wind velocity in the $x$-direction, and similarly
$V$ and $W$; $\Theta=\mu\theta$, where $\theta=\theta\left(  x,y,\eta\right)
$ is the potential temperature; $\phi=\phi\left(  x,y,\eta\right)
=\overline{\phi}+\phi^{\prime}$ is the geopotential; and $Q_{m}=\mu q_{m}$ is
the moisture contents of the air. The variables in the state $\Phi$ evolved by
(\ref{eq:WRF}) are called prognostic variables. Other variables computed from
them, such as the hydrostatic pressure $p$, the thermodynamic temperature $T$,
and the height $z$, are called diagnostic variables. The variables that
contain $\mu$ are called coupled. The value of the right-hand side $R\left(
\Phi\right)  $ is called tendency. See \cite[pp. 7-13]{Skamarock-2008-DAR} for
details and the form of $R$.

The system (\ref{eq:WRF}) is discretized in time by the explicit 3rd order
Runge-Kutta method%
\begin{align}
\Phi_{1} &  =\Phi^{t}+\frac{\Delta t}{3}R\left(  \Phi^{t}\right)  \nonumber\\
\Phi_{2} &  =\Phi^{t}+\frac{\Delta t}{2}R\left(  \Phi_{1}\right)  \nonumber\\
\Phi^{t+\Delta t} &  =\Phi^{t}+\Delta tR\left(  \Phi_{2}\right)
\label{eq:rk3}%
\end{align}
where only the third Runge-Kutta step includes tendencies from physics
packages, such as the fire module \cite[p. 16]{Skamarock-2008-DAR}. In order
to avoid small time steps, the tendency in the third Runge-Kutta step also
includes the effect of substeps to integrate acoustic modes.

\subsection{Surface schemes}

In real cases, non zero \texttt{sf\_sfclay\_physics} should be selected to
enable the surface model, allowing for proper interaction between the
atmosphere and the land surface. In idealized cases, users have an option of
the basic surface initialization, intended to be used without the surface
model, or the full surface initialization (\texttt{sfc\_full\_init=1}). The
latter allows for using all standard land surface models even in idealized
cases. For idealized cases with full surface initialization, land surface
properties like roughness length, albedo etc., are defined through the land
use category. The surface scheme utilizes a gridded array containing the
number of landuse category, defined in a text file (\texttt{LANDUSE.TBL}),
which specifies the roughness length and other surface properties, both for
the real and idealized cases. The land use categories may be also defined
directly trough the namelist variables or read in from an external file
containing a 2D landuse matrix (see also Sec.~\ref{sec:data-input}).

\section{Coupling of the fire and the atmospheric models}

\label{sec:coupling}

The terrain gradient is computed from the terrain height at the best available
resolution and interpolated to the fire mesh in preprocessing. Interpolating
the height and then computing the gradient would cause jumps in the gradient,
which affect fire propagation, unless high-order interpolation is used.

In each time step of the atmospheric model, the fire module is called from the
third step of the Runge-Kutta method. First the wind is interpolated to a
fixed height $z_{f}$ above the terrain (currently, 6.1m following BEHAVE),
assuming the wind log profile%
\[
u\left(  z\right)  \approx\left\{
\begin{array}
[c]{cc}%
\operatorname*{const}\log\frac{z}{z_{0}}, & z\geq z_{0},\\
0 & 0\leq z\leq z_{0},
\end{array}
\right.
\]
where $z$ is the height above the terrain and $z_{0}$ is the roughness height.
For a fixed horizonal location, denote by $z_{1},$ $z_{2},\ldots$ the heights
of the centers of the atmospheric mesh cells; these are computed from the
geopotential $\phi$, which is a part of the solution. The horizontal wind
component $U\left(  z_{f}\right)  $ is then found by interpolating the values
$U\left(  z_{0}\right)  =0,U\left(  z_{1}\right)  ,U\left(  z_{2}\right)
,...$ from the nodes $\log z_{0}$, $\log z_{1},\log z_{2},\ldots$ to $\log
z_{f}$; if $z_{f}\leq z_{0}$, we set $U\left(  z_{f}\right)  =0$. The
$V$-component of the wind is interpolated in the same way.

The fire model then makes one time step:

\begin{enumerate}
\item If there are any active ignitions, the level-set function is updated and
the ignition times of any newly ignited nodes are set following Sec.
\ref{sec:ignition}.

\item The numerical scheme (\ref{eq:Heun})-(\ref{eq:Godunov}) for the level
set equation (\ref{eq:levelset}) is advanced to the next time step.

\item The time of ignition set for any any nodes that were ignited during the
time step, from (\ref{eq:set-ignition-time}).

\item The fuel fraction is updated following Sec.~\ref{sec:fuel-fraction}.

\item The sensible and latent heat flux densities are computed from
(\ref{eq:sensible-heat-flux}) and (\ref{eq:latent-heat-flux}) in each fire
model cell.

\end{enumerate}
The resulting heat flux densities are averaged over the fire cells that
make up one atmosphere model cell and inserted into the atmospheric model,
which then completes its own time step.
The heat fluxes from the fire are inserted into the atmospheric model as
forcing terms in the differential equations of the atmospheric model into a
layer above the surface, with assumed exponential decay with altitude. Such a
scheme is needed because WRF does not support flux boundary conditions. The
sensible heat flux is inserted as the tendency of the potential temperature
$\theta$, equal to the vertical divergence of the heat flux,%
\[
\frac{d\left(  \mu\theta\right)  }{dt}\left(  x,y,z\right)  =R_{\Theta}\left(
\Phi\right)  +\frac{\mu\left(  x,y\right)  \phi_{h}\left(  x,y\right)
}{\sigma\varrho\left(  x,y,z\right)  }\frac{\partial}{\partial z}\exp\left(
-\frac{z}{z_{ext}}\right)  ,
\]
where $R_{\Theta}\left(  \Phi\right)  $ is the component of the tendency in
the atmospheric model equations (\ref{eq:WRF}), $\sigma$ is the specific heat
of the air, $\varrho\left(  x,y,z\right)  $ is the density, and $z_{ext}$ is
the heat extinction depth, a parameter. The latent heat flux is inserted
similarly into the tendency of the vapor concentration $q_{m}$ by%
\[
\frac{d\left(  \mu q_{m}\right)  }{dt}\left(  x,y,z\right)  =R_{Q_{m}}\left(
\Phi\right)  +\frac{\mu\left(  x,y\right)  \phi_{q}\left(  x,y\right)
}{L\varrho\left(  x,y,z\right)  }\frac{\partial}{\partial z}\exp\left(
-\frac{z}{z_{ext}}\right)  ,
\]
where $L$ is the specific latent heat of the air.

\section{Software structure}

\label{sec:software}

\subsection{Grids and parallel execution}

Parallel computing imposes a significant constraint on user programming
technique. WRF uses the RSL parallel infrastructure \cite{Michalakes-1999-RPR}%
. RSL divides the domain horizontally into patches. Each patch executes in a
separate MPI\ process and is further divided into tiles, which execute in
separate OpenMP threads. Communication between tiles is accomplished at the
end of OpenMP parallel loop over tiles. The fire grid has refined tiles in the
same location as atmospheric grid tiles. The patches are declared in memory
with larger bounds than the patch size, and communication between patches is
accomplished by HALO\ calls (actually, includes of generated code), which
update a layer of nodes beyond the patch boundary from other patches. The fire
module computational code itself is designed to be \emph{tile-callable}; it
executes on a single tile, assuming it can safely read data from a layer of
nodes beyond the tile boundary. The communication (OpenMP loops or
HALO\ calls) occurs outside the computational routines; this means that
whenever communication is necessary, the fire module must exit, and then
continue from the correct code location on the next call.

\subsection{Software layers}

\begin{figure}[pt]
\begin{center}
\fboxsep=0.1in \fbox{\includegraphics[width=4in]{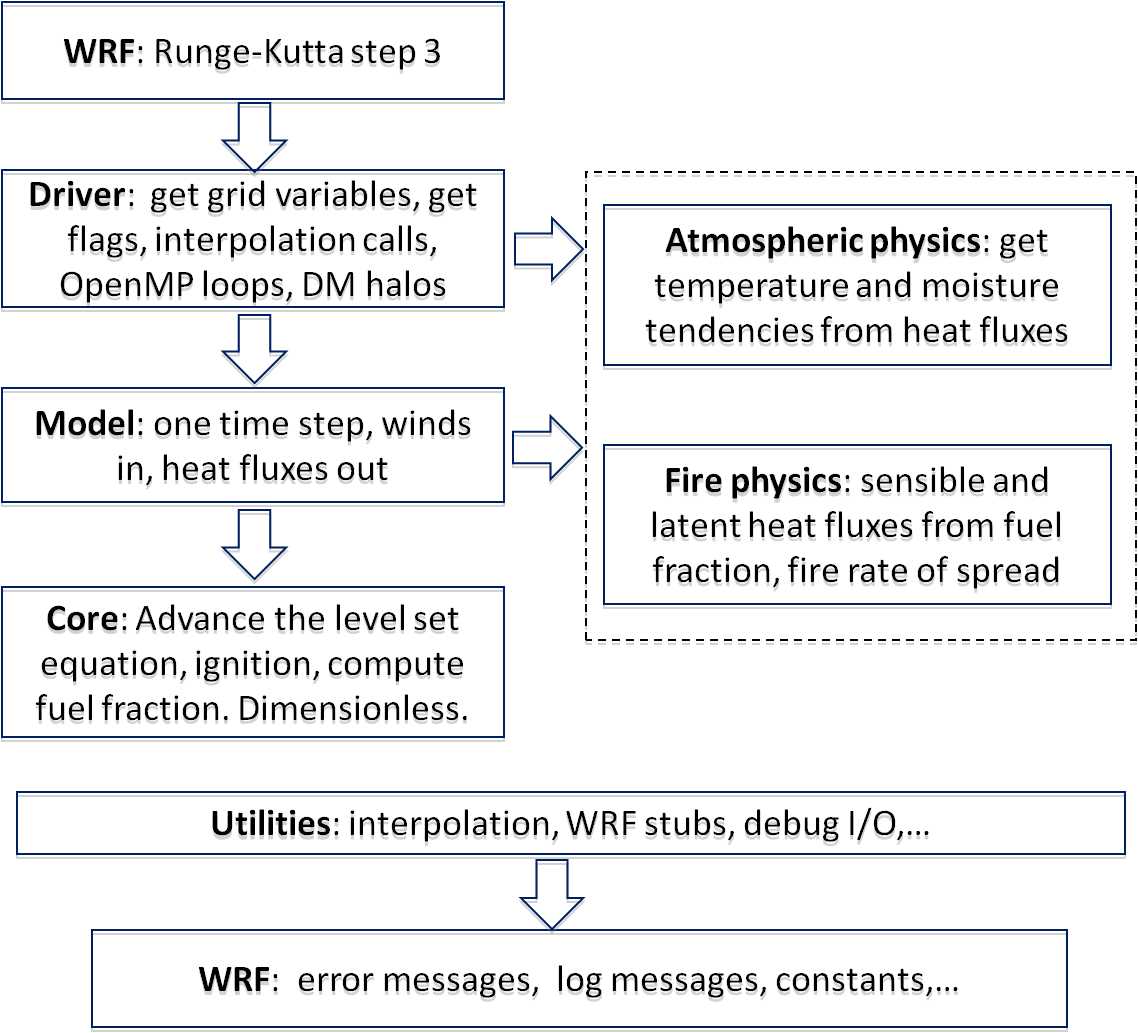}}
\end{center}
\caption{Structure of WRF-Fire software. All physics routines are in the
dashed box. The Utilities layer is used by all other components above it.}%
\label{fig:structure}%
\end{figure}

The fire module software is organized in several isolated layers
(Fig.~\ref{fig:structure}). The \emph{driver layer} serves to interpolate and
otherwise translate the variables between WRF and the fire module, and it
contains all exchange of data between the tiles in parallel execution. The
rest of the code executes on a single tile, assuming that the needed values
from neighboring tiles are already present. This structure is needed so that
the rest of the code can conform to the WRF coding conventions
\cite{WRF-conventions}. Only the driver layer depends on WRF; the rest of the
fire module can be used as standalone code, independent of WRF. WRF
infrastructure is accessed only through stubs in the \emph{utility layer }so
that it can be easily emulated in the standalone code. The \emph{model layer}
is the entry point to the fire module. The \emph{core layer} is the engine of
the fire model, described in Sec.~\ref{sec:core}. The \emph{fire physics
layer} evaluates the fire spread rate and heat fluxes from fuel properties,
and the \emph{atmospheric physics} layer mediates the insertion of the fire
fluxes into the atmosphere, as described in Sec.~\ref{sec:coupling}. One of
the goals of the design is that the only components that need to be modified
when the fire module is connected to another atmospheric model in future are
the driver layer, the atmospheric physics layer, and WRF stubs in the utility layer.

\section{Data input}

\label{sec:data-input}

WRF ideal run is used for simulations on artificial data. An additional
executable, \texttt{ideal.exe}, is run first to create the WRF input. A
different \texttt{ideal.exe} is built for each ideal case, and the user is
expected to modify the source of such ideal case to run custom experiments.
The ideal run for fire supports optional input of gridded arrays for land
properties, such as terrain height and roughness height. This allows a user to
execute simulations that go beyond what would normally be considered an ideal
run and simplifies custom data input; the simulation of the FireFlux
experiment was done in this way \cite{Kochanski-2010-EFP}.

In a real run, the WRF Preprocessing System (WPS) \cite[Chapter 3]%
{Wang-2010-AUG} takes meteorological and land-use data in a number of commonly
used formats and prepares it for WRF to use as initial and boundary
conditions. WPS has been extended to process fine-scale land data for use with
the fire model, such as topography and fuel.

\section{Acknowledgements}

The interpolation of ignition time in Sec.~\ref{sec:fuel-fraction} was
implemented by Volodymyr Kondratenko, who, with Minjeong Kim, has also helped
to implement the analytic integration of the fuel fraction in
Sec.~\ref{sec:fuel-fraction}. John Michalakes developed the support for the
refined surface fire grid in WRF. The devepers would like to thank Ned Patton
for providing a copy of his prototype code, and Janice Coen for providing a
copy of CAWFE. Other contributions to the model are acknowledged by
bibliographic citations in the text. This research was supported by NSF grant AGS-0835579
and by NIST Fire Research Grants Program grant 60NANB7D6144.

\bibliographystyle{ametsoc}
\bibliography{../../references/geo,../../references/other}

\end{document}